\documentclass[journal=jacsat,manuscript=article]{achemso}

\usepackage[version=3]{mhchem} 
\usepackage{booktabs}
\usepackage[table,xcdraw]{xcolor}
\usepackage{dcolumn}
\usepackage{amssymb}
\usepackage{subfig}
\usepackage{tabularx}
\usepackage{footnote}
\usepackage{threeparttable}
\usepackage{xcolor}

\usepackage{lipsum} 

\usepackage{tikz,xcolor,hyperref}

\definecolor{lime}{HTML}{A6CE39}
\DeclareRobustCommand{\orcidicon}{%
	\begin{tikzpicture}
	\draw[lime, fill=lime] (0,0) 
	circle [radius=0.16] 
	node[white] {{\fontfamily{qag}\selectfont \tiny ID}};
	\draw[white, fill=white] (-0.0625,0.095) 
	circle [radius=0.007];
	\end{tikzpicture}
	\hspace{-2mm}
}

\foreach \x in {A, ..., Z}{%
	\expandafter\xdef\csname orcid\x\endcsname{\noexpand\href{https://orcid.org/\csname orcidauthor\x\endcsname}{\noexpand\orcidicon}}
}

\author{Domna G. Kotsifaki}
\author{Viet Giang Truong}
\author{S\'{i}le Nic Chormaic}
\email{sile.nicchormaic@oist.jp}

\affiliation{Light-Matter Interactions for Quantum Technologies Unit, Okinawa Institute of Science and Technology Graduate University, Onna-son, Okinawa, 904-0495, Japan}

\title[An \textsf{achemso} demo]
  {Fano-Resonant, Asymmetric, Metamaterial-Assisted Tweezers for Single Nanoparticle Trapping}

\begin{document}

\begin{abstract}
Plasmonic nanostructures can overcome Abbe's diffraction limit to generate strong gradient fields, enabling efficient optical trapping of nano-sized particles. However, it remains challenging to achieve stable trapping with low incident laser intensity. Here, we demonstrate a Fano resonance-assisted plasmonic optical tweezers (FAPOT), for single nanoparticle trapping in an array of asymmetrical split nano-apertures, milled on a 50~nm gold thin film. Stable trapping is achieved by tuning the trapping wavelength and varying the incident trapping laser intensity. A very large normalized trap stiffness of 8.65 fN/nm/mW for 20 nm polystyrene particles at a near-resonance trapping wavelength of 930 nm was achieved. We show that trap stiffness on resonance is enhanced by a factor of 63 compared to off-resonance conditions. This can be attributed to the ultra-small mode volume, which enables large near-field strengths and a cavity Purcell effect contribution. These results should facilitate strong trapping with low incident trapping laser intensity, thereby providing new options for studying transition paths of single molecules, such as proteins, DNA, or viruses.
\end{abstract}


\section{Introduction}

Metallic nanostructures have  attracted considerable attention for enhancing light-matter interactions due to  unique properties that enable them to concentrate light beyond the diffraction limit, thereby enabling strong field confinement\cite{R71,R59}. By patterning metal nanostructures in arrays\cite{R72,R59}, one can achieve lower radiation losses, higher quality factors, and large field enhancements over wide effective cross-sections, characteristics that are important for a variety of projected applications. Various nanostructure arrays have been fabricated to achieve ultrasensitive biodetection\cite{R73,R61}, to create advanced liquid crystal devices\cite{R62}, to crystallize proteins\cite{R63}, to deliver drugs\cite{R68}, to monitor cancer\cite{R67} and to perform biomedical imaging\cite{R64}. 

Likewise, during the past few years, there has been growing interest in precise manipulation of nano-sized objects. Plasmonic optical tweezers (POTs)\cite{R74,R75,R41,R91,R87} provide a label-free means of single-nanoparticle characterization\cite{R11,R14,R17,R27,R78,R1} and contribute to the development of lab-on-chip analytical platforms. Typically, POTs rely on evanescent waves around metallic nanostructures, which produce localized intensities that enhance optical forces at nanoscale. For example, Roxworthy et al.\cite{R25} introduced an array trapping platform with the ability to control single-particle and multi-particle trapping by adjusting the trapping incident power. The authors claim that their platform opens the door for specific particle trapping selectivity based upon size, mass, and/or refractive index. However, in some cases where manipulation of temperature-sensitive bioparticles is needed, nano-aperture based POTs, which enable stable trapping with low incident intensity are introduced\cite{R77,R11,R17,R27,R36,R69}, thereby minimizing phototoxicity\cite{R69}. By arraying nano-aperture units\cite{R11,R25,R86,R88} in metallic substrates, many trapping sites can be activated at the same time, permitting simultaneous analysis of several particles locally trapped in well-defined positions of the plasmonic nanostructure and providing an alternative method of particle crystallization. Therefore, selectivity of biomolecule trapping in a heterogeneous environment could have a significant impact on defining structural information during transition paths\cite{R4}, leading among other things, to more precise drug design\cite{R39,R40}.   

Currently, metamaterials are being widely studied owing to their exceptional properties, which originate from structuring on a sub-wavelength scale\cite{R21,R16,R30,R34}. In particular, they pave the way for a new world of remarkable applications, including slow light devices\cite{R46}, cloaking\cite{R47}, and lasing spasers\cite{R49}. Fano-resonant, asymmetric, split-ring metamaterials exhibit sharp resonances caused by interference between super-radiant and subradiant plasmonic modes\cite{R9,R53}. By exploiting Fano resonance, it is possible to confine light more efficiently, characterized by a steeper dispersion than the dipole resonance. This makes metamaterials promising for nanoscale implementation. Thanks to their extreme sensitivity to local geometrical changes, they have been proposed as a quantitative biosensing platform where the Fano resonance shows sensitivity for protein recognition\cite{R34}. Additionally, a Fano metamaterial consisting of a disc-double, split-ring resonator enhances the chiral gradient force on sub-10-nm enantiomers, enabling new options for chiral trapping and separation\cite{R50}. The fact that the Fano resonance peak is highly sensitive to modifications of its surrounding medium\cite{R16} renders metamaterials particularly attractive for stable optical trapping applications.
 
In this work, we introduce a novel approach for strong, single-nanoparticle trapping, based on Fano resonance-assisted plasmonic optical tweezers (FAPOT). To the best of our knowledge, this is the first experimental demonstration of a FAPOT metamaterial for nanoparticle trapping. This approach provides superior stable plasmonic trapping for nanoscale particles, demonstrating enhanced trapping performance with near-infrared wavelength tunability. We fabricated an array of asymmetric, split-ring (ASR) plasmonic nanostructures that were used as trapping substrates. We studied the trapping performance via power- and wavelength-dependent characterization, revealing the highest effective trap stiffness reported  to date for 20-nm polystyrene nanoparticles, equal to 8.65 fN/nm/mW under resonance conditions. We experimentally investigated polarization-dependent properties of the metamaterial via trap stiffness measurements. Owing to metamaterial properties, enhancement of the trap stiffness could indicate the cavity quantum electrodynamic effect \cite{R89}. We conclude that the ability of our system/platform to perform steady and dynamic optical manipulation makes a variety of  lab-on-chip applications possible.

\section{Experimental Details}

\subsection{Sample Fabrication} 
An array of asymmetric split-rings (ASRs) was fabricated using focused ion beam (FIB-FEI Helios G3UC) milling on a 50~nm gold thin film (PHASIS, Geneva, BioNano) at 30 kV energy and 2 pA beam current. The metamaterial device, which was used for trapping experiments, consisted of 17 ($\it{x}$-direction) and 15 ($\it{y}$-direction) identical ASR units, \textit{i.e.}, metamolecules. Figure 1a shows a scanning electron microscope (SEM) image of 4x4 ASR metamolecules at an angle of $52^{\circ}$ from the surface normal. 

\begin{center}
 \begin{figure}
\centering
\includegraphics[width = 185 mm, height = 90 mm]{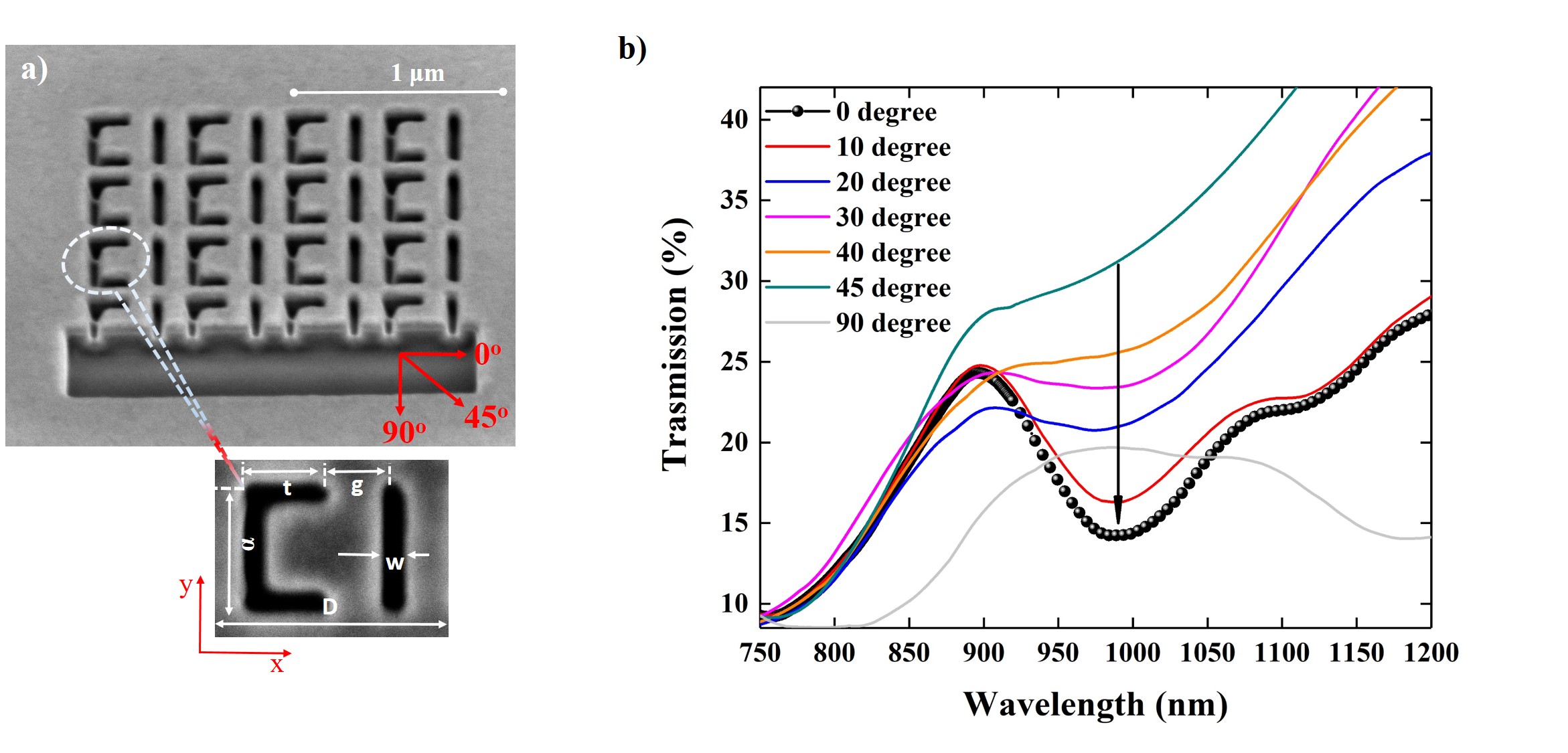}
\caption{(a) Scanning electron microscope (SEM) image, viewed at $52^{\circ}$ from the surface normal. The metamaterial structure is a 4x4 array of ASR metamolecules in a 50~nm gold thin film. Zoom: Metamolecule unit with feature dimensions: D = 400.0$\pm 2.1$ nm, vertical slit $\alpha$ = 310.0$\pm1.3$ nm,  horizontal slit t = 164.4$\pm2.6$ nm, gap g = 101.3$\pm 3.2$ nm and slit width w = 44.3$\pm 1.8$ nm. (b) Transmission spectra featuring the resonance of an array of 36 metamolecules (6x6 units) of the metamaterial in de-ionized water versus the polarization angle of the microspectrophotometer light source. The arrow indicates resonance changes.} 
\label{Figure1}
 \end{figure}  
\end{center}


\subsection{Experimental Setup}

The experimental set-up used in this work (Figure S1 in the Supporting Information-SI) is based on an inverted microscope. The optical tweezers consist of a tunable, continuous-wave (cw) Ti:Sapphire laser focused using a 1.3x numerical aperture (NA) oil immersion objective lens (OLYMPUS UPlanFL N 100x) on the metamaterial device, which was mounted on an xyz-piezo nanopositioning stage. From recorded images, we estimated that the laser spot size at the objective focus was approximately 2.0 $\mu$m and the number of units that could be illuminated on the array was 5x5, \textit{i.e.}, 25 metamolecules. According to Fedotov et al.\cite{R8} the quality factor, \textit{Q} (see SI), of the ASR metamaterial strongly depends on the size of the illuminated area, \textit{i.e.}, the total number of ASR units involved in the interaction with the incident light. Thus, the more ASR units are exposed to the incident laser beam, the more the material losses (\textit{i.e.}, Ohmic losses in the metal and dissipation losses in the glass substrate) and the radiation losses are reduced\cite{R8}. As a result, the number of  ASR units illuminated by the trapping laser beam ensures a high \textit{Q}-factor of the metamaterial\cite{R8} and a narrow Fano resonance peak under resonance conditions.

We controlled both the polarization and incident power of the trapping beam, which was limited to a maximum power of 10 mW at the sample plane. Detection of the trapping event was accomplished by collecting the transmitted laser light through a 50x objective lens (Nikon CF Plan) and sending it to an avalanche photodiode (APD430A/M, Thorlabs). A second APD was used to collect reflected laser light through an oil immersion objective lens. APD signals were recorded using a data acquisition board (DAQ) at a frequency of 100 kHz with LabVIEW. In order to calculate trap stiffness, we only analyzed data for which the APD signal of the reflected laser light was in the opposite direction (step-like jump up or down) to the APD signal of the transmitted laser light (step-like jump down or up). This ensured that the step-like jump was due to an optical trapping event.

The metamaterial device was attached to a cover glass with adhesive microscope spacers (Grace BioLabs, Sigma-Aldrich, GBL654002), forming a microwell. The microwell contained polysterene (PS) particles with a mean diameter of 20 nm (ThermoFisher Scientific, F8786) in heavy water (D$_{2}$O) with a 0.0625\% mass concentration. A small amount of surfactant (Detergent Tween 20 with 0.1\% volume concentration) was used to minimize aggregation. The microwell was mounted and fixed on top of a piezoelectric translation stage.   

\section{Results}

Transmission spectra of the metamaterial were obtained using a microspectrophotometer (MCRAIC 20/30 PV). Figure 1b shows the transmission spectra of the ASR metamaterial in de-ionized water as a function of the polarization angle. By careful design of the ASR metamolecules, the broad super-radiant and narrow subradiant plasmon modes\cite{R16,R23} can overlap so as to provide an asymmetric Fano line that appears at ~892 nm (Figure 1b). For vertical polarization (\textit{i.e.,} $90^{\circ}$ gray line in Figure 1b), the metamaterial devices did not show spectral features originating from asymmetrical structuring. Note that the transmission spectrum difference between the two orthogonal polarizations (Figure 1b), implies that this metamaterial can be used to improve detection sensitivity for horizontal polarization, \textit{i.e.}, at $0^{\circ}$ . 

Figure 2 shows a raw data trace of the APD signal versus time. The trapping wavelength was 910 nm and the incident laser power after the objective lens was 3.0 mW, \textit{i.e.}, resulting in a trapping laser intensity of 0.96 mW/$\mu$m$^{2}$ at the sample plane. When a PS particle was approaching excited ASRs, the optical gradient force pulled the particle toward the nano-aperture of the ASR metamaterial. Figure 2a shows the cycle of trapping and releasing of a 20-nm PS particle in an array of ASRs. As the PS particle was trapped, APD signals were recorded and, subsequently, the laser was blocked to release the trapped particle from the nanostructure. After 7 sec, we re-illuminated the metamaterial, leading to another step-like signal showing nearly identical trap and release intensity levels. This reversible trap/release cycle was repeated several times demonstrating the high repeatability of the trapping process. Figure 2b shows a single step-like jump in the APD signal, typically within 1 min, which we attribute to a nano-aperture-based trapping event. Fluctuations in the APD signal correspond to the Brownian motion of the particle in the trap. By changing the trapping laser power and the trapping time duration, we observed multiple nanoparticle trapping events (Figure S2-SI). However, in this work, in order to determine the trap stiffness enhancement factor, we only analyze data for the first observable single-nanoparticle trapping event.

\begin{figure}[ht!]
\centering
\includegraphics[width = 175 mm, height = 140 mm]{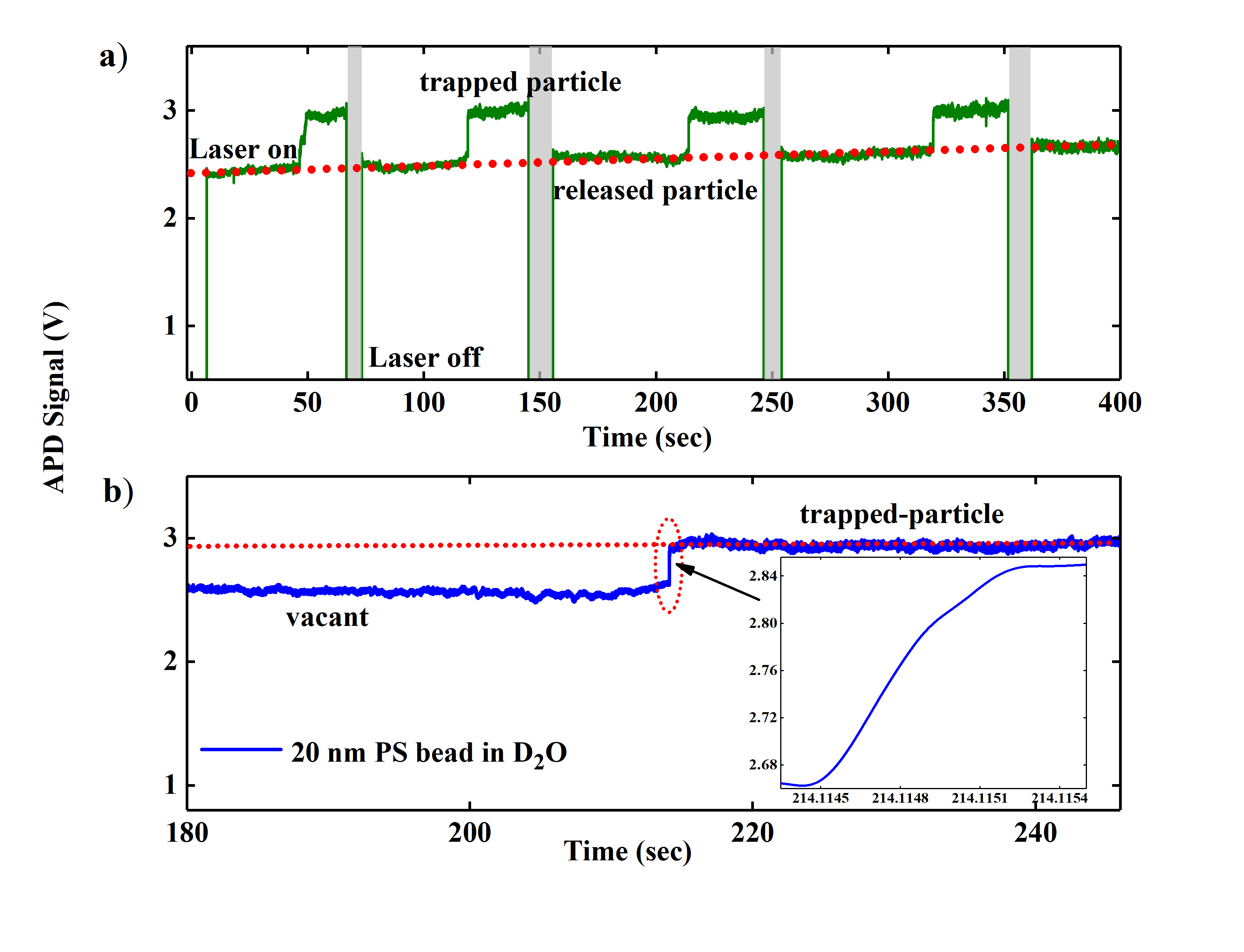}
\caption{(a) Time trace of the APD signal through the ASR metamaterial in heavy water containing suspended 20-nm PS particles as a function of the trapping laser beam for a trapping laser wavelength of 910 nm. Gray regions indicate times when the laser was blocked, triggering release of the trapped PS particle. (b) Time trace of a typical trapping event of a 20-nm PS particle. Inset: a zoom of the trapping event at 214.1 sec, representing a time interval of 0.9 ms. The trapping laser intensity at the sample plane was 0.96 mW/$\mu$m$^{2}$.}
\label{figure2}
 \end{figure}

For small displacements (\textit{i.e.}$, |x|$$\ll$ $\lambda$, where $\lambda$ is the trapping wavelength)\cite{R17} of the particle from the trap, 
the overdamped Langevin equation\cite{R17} can be applied:
\begin{equation}
    \gamma \dot{x}(t)+\kappa_{tot} x(t)=\xi(t),
\end{equation}
where \textit{$\gamma$} is the viscous damping, \textit{x(t)} the displacement of the particle from the equilibrium point, \textit{$\kappa_{tot}$} the total trap stiffness, which is proportional to the trapping laser intensity, and \textit{$\xi$(t)} is the thermal fluctuation. Using an exponential fit of the trapping transient region, trap stiffness, $\kappa_{tot}$, is determined from the following equation\cite{R14}:
\begin{equation}
    \tau = \frac{\gamma}{\kappa_{tot}},
\end{equation}
where \textit{$\tau$} is the exponential decay time. For the trap stiffness calculation, the effective distance between the 20-nm PS particle and the surface of the ASR wall is 5 nm\cite{R14,R11}. This is due to the surface roughness of the fabricated ASR device. We consider the Stokes' drag coefficient and adjust it using Faxen's correction factor for the trap stiffness calculations (see SI-S3). Thence, the trap stiffness increases by a factor of 1.78 \cite{R11,R14} (see Eq.S1-SI).
 

 Figure 3 shows the trap stiffness as a function of the trapping laser intensity ranging from 0.23 mW/$\mu$m$^{2}$ to 1.4 mW/$\mu$m$^{2}$, for trapping of a 20-nm PS particle. As expected, the linear dependence of trapping stiffness versus laser intensity is obvious at both 910 nm and 950 nm, with deviation from linearity as the trapping wavelength approaches the resonance peak at 930 nm. We obtained a maximum trap stiffness, \textit{$\kappa_{tot}$} = 1.99 fN/nm, for the lowest laser intensity, \textit{i.e.}, 0.23 mW/$\mu$m$^{2}$, corresponding to a normalized value of 8.65 fN/nm/mW for a trapping laser intensity of 1 mW/$\mu$m$^{2}$.  This normalized trap stiffness is the highest experimental value reported to date for trapping of dielectric nanoparticles. Enhancement of trapping stiffness is a feature of the Fano resonant effect of the metamaterial on the FAPOT trapping. 
To further investigate the origin of trap stiffness enhancement, we studied the dependence of trap stiffness on the trapping wavelength. 

\begin{figure}
\centering
\includegraphics[width = 115 mm, height =170 mm]{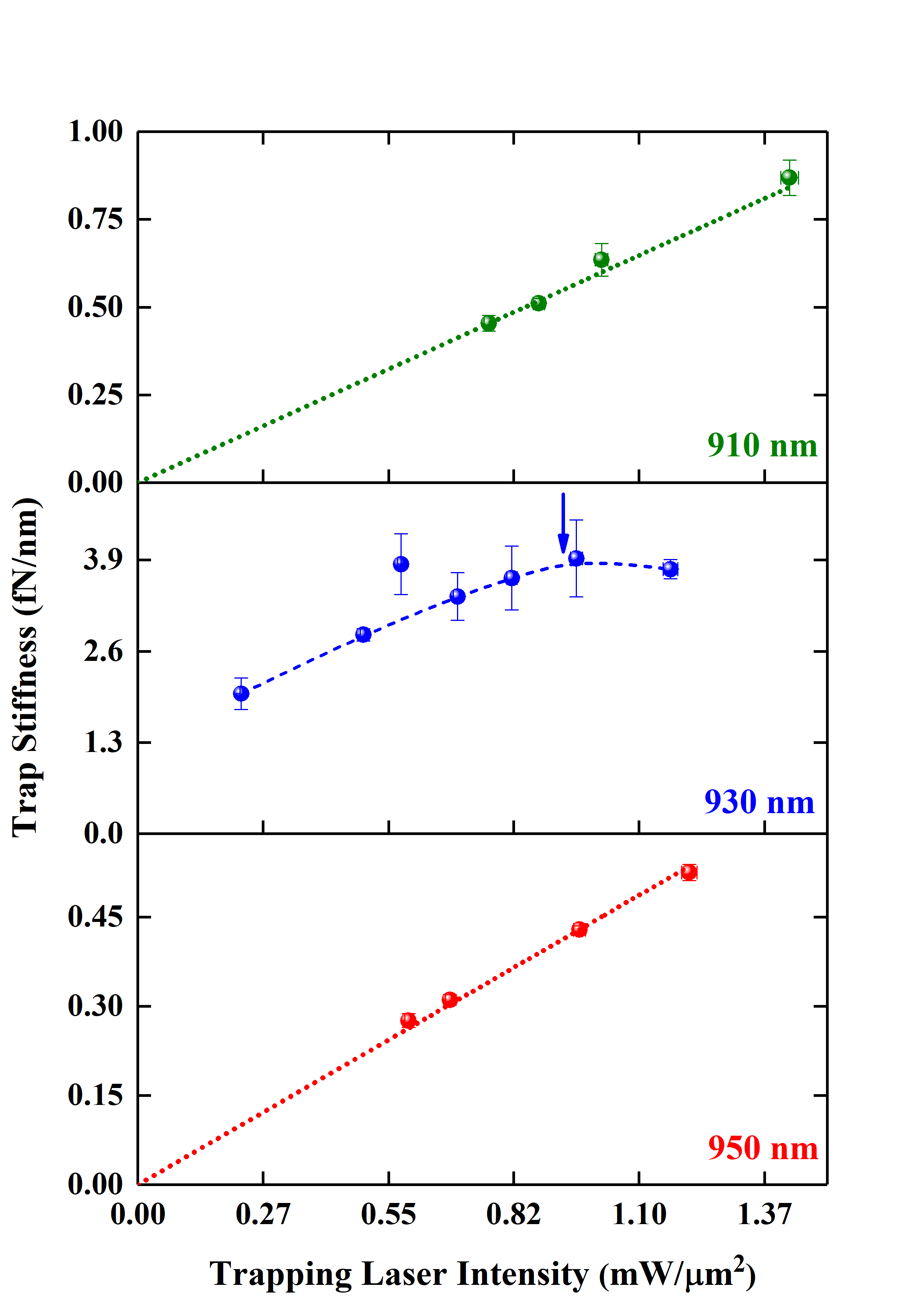}
\caption{Trap  stiffness  of  a  single  20-nm  PS  particle  as  a  function  of  incident  laser intensity for laser trapping wavelengths of 910 nm, 930 nm, and 950 nm. Red and green dashed lines: Linear fit to the data obtained using the ASR metamaterial. The red dashed line has a slope of 0.16 [(fN/nm)/(mW/$\mu$m$^{2}$)] and the green dashed line has a slope of 0.08 [(fN/nm)/(mW/$\mu$m$^{2}$)]. The blue arrow indicates the point at which the data deviate from a linear trend. The y-error corresponds to the standard deviation of the trap stiffness measurements.}
\label{Figure3}
 \end{figure}
\begin{figure}[ht!]
\centering
\includegraphics[width = 135 mm, height =180 mm]{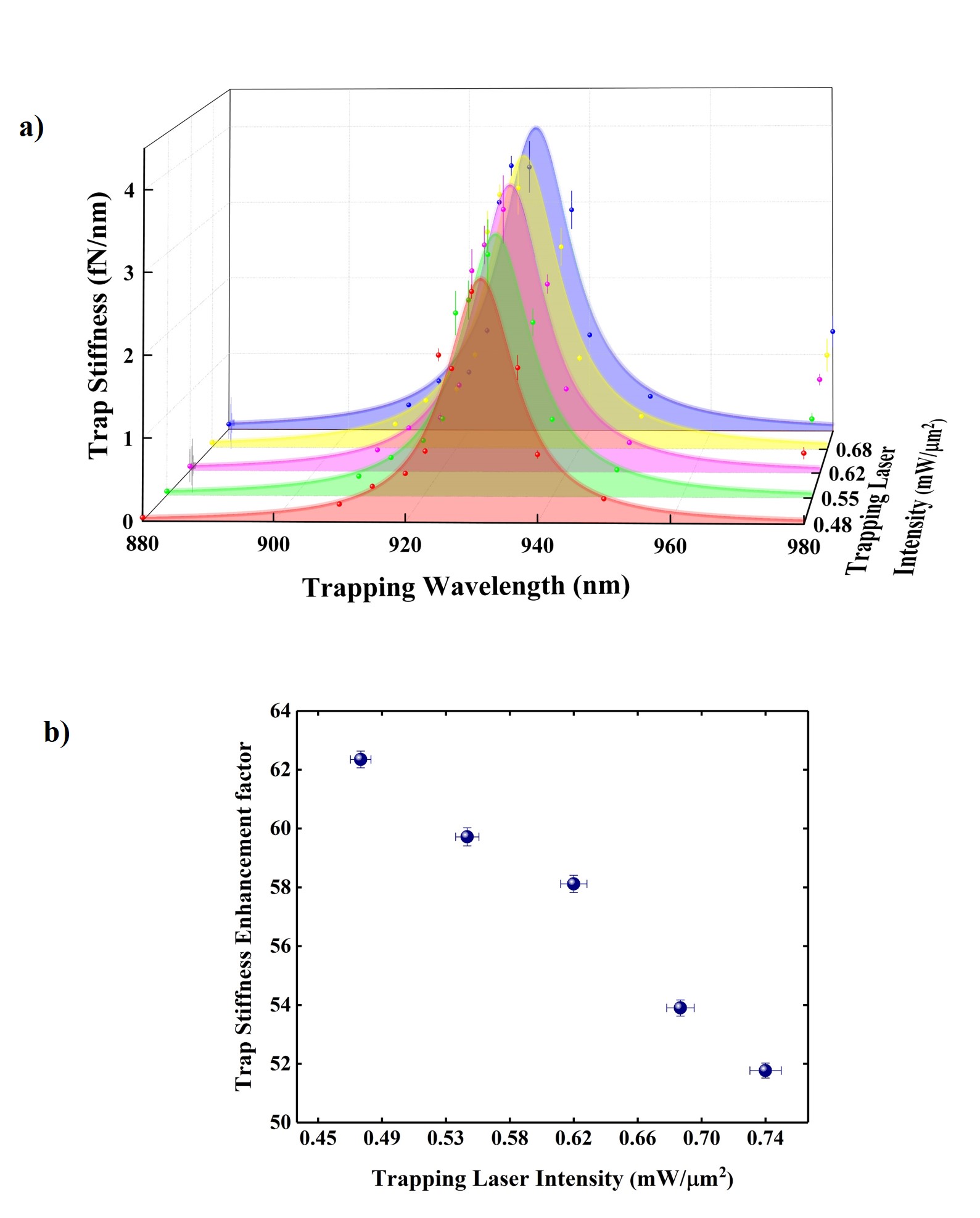}
\caption{(a) Trap  stiffness  of  a  single 20-nm PS  particle as a function of trap intensity and trapping wavelength. The y-error corresponds to the standard deviation of the trap stiffness measurements.(b) Trap stiffness enhancement factor as a function of incident trapping laser intensity 
Enhancement is normalized to the trap stiffness measured at 880 nm, far from the resonance peak. The $x$-error corresponds to the standard deviation of the power measurements and the $y$-error to the propagation error of the trap stiffness measurements.
}
\label{Figure4}
 \end{figure}

Figure 4a shows the trap stiffness as a function of trapping laser intensity and the trapping wavelength. For each intensity value, data are fitted with a Lorentz function, represented by the solid curve. For all fitting curves, we determined the existence of a resonance wavelength at 931.36 $\pm$ 1.76~nm, where the trap stiffness is maximal. 
We noticed that the resonance peak, calculated by fitting the data of Figure 4a, is red-shifted when compared to the transmission spectra in Figure 1b for a polarization angle of $0^\circ$ (black line in  Figure 1b). In order to investigate the nature of this observation, we perform transmission spectrum measurements of the ASR metamaterial in air, in which we estimate a resonance at 872 nm (data  not presented). From  Figure 1b, the resonance peak of the ASR metamaterial in de-ionized water is 892 nm. We plot the resonance peaks as a function of the refractive index, and using a linear fit, we evaluate the resonance for a polystyrene-coated ASR metamaterial around 907 nm (we assume that the refractive index of polystyrene is 1.58). When a particle is trapped in a nanoscale cavity, leading to enhancement of the group refractive index of the local environment around the trapping site, the resonance frequency of the cavity depends strongly on the local group index and the particle position\cite{R17}. As a result, the particle itself causes a red shift toward 907 nm, the magnitude of which is determined by the combination of the cavity quality factor, \textit{Q}, and the ratio of the particle volume to the cavity mode volume\cite{R17}. We noticed that trap stiffness measurements indicate a resonance at 931 nm. Given that this wavelength is longer than 907 nm, we assume that this observation is due to the nanoparticle's dynamic role contribution \cite{R88}.

In Figure 4b, we plot the trap stiffness enhancement factor as a function of trapping laser intensity. The enhancement factor is normalized to the trap stiffness value obtained at 880 nm, far from the Fano resonance peak. Interestingly, we note that a higher trap stiffness enhancement is obtained for low trapping laser intensities and it diminishes as trapping intensity increases, indicating that the ASR metamaterial device is very sensitive to trapping laser intensity changes.

\begin{figure}[ht!]
\centering
\includegraphics[width = 175 mm, height =145 mm]{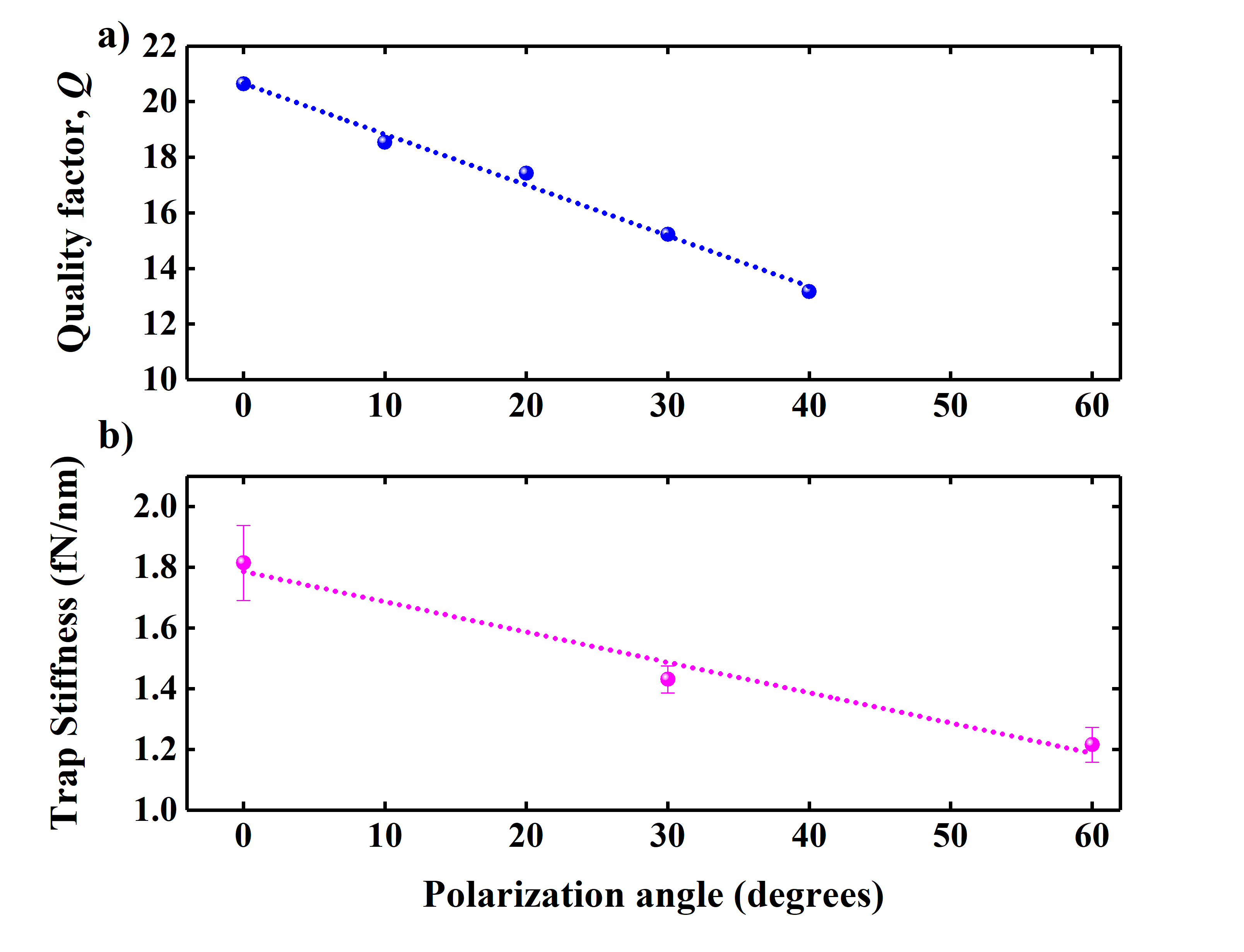}
\caption{(a) Quality factor, \textit{Q}, as a function of the polarization angle of the microspectrophotometer light source. The dashed line indicates the linear fit to the \textit{Q}-factor data and has a slope of 0.18. (b) Trap stiffness for a single 20-nm PS particle as a function of polarization angle for a trapping wavelength of 937 nm and a trapping intensity of 0.82 mW/$\mu$m$^{2}$ at the sample position. The $y$-error corresponds to the standard deviation of the trap stiffness measurements. The dashed line indicates the linear fit to the trapping data and has a slope of 0.010. 
}
\label{Figure5}
 \end{figure}
 
Figure 5a is a plot of the quality factor, \textit{Q}, of the ASR metamaterial, as a function of polarization angle of the microspectrophotometer light source. The $Q$-factor was calculated using Eq. S2 (see SI) based on data from Figure 1b. We could not calculate the \textit{Q}-factor for angles larger than $40^{\circ}$. We observe that the \textit{Q}-factor of the ASR metamaterial decreases as a function of polarization angle. This may imply that the net magnetic flux of the ASR metamaterial\cite{R8} vanishes due to non-efficient excitation of the resonance mode. Particularly, by changing the polarization of the incident light, interactions between individual ASR units mediated by the magneto-induced surface waves\cite{R8}, which contribute to excitation of the Fano resonance peak, are less strong, leading to an increase of scattering losses through magnetic dipole radiation\cite{R8}. Owing to the above-mentioned scattering loss analysis, we assume that the \textit{Q}-factor diminishes as the polarization angle changes.  
Moreover, rotating the polarization reveals the difference in the near-field potential landscape experienced by the trapped PS particle, leading to a variation in the magnitude of the trap stiffness (Figure 5b). For a horizontal polarization  (\textit{x}-direction or $0^{\circ}$  in Figure 1a), Fano resonance dominates, whereas it vanishes for vertical polarization (\textit{y}-direction or $90^{\circ}$ in Figure 1a). The fact that trap stiffness follows a similar trend as the \textit{Q}-factor, confirms the exceptional polarization-dependent properties of the ASR metamaterial, which modify the magnitude of trap stiffness.   

\newpage
\section{Discussion}

In Table 1, we compare our experimentally measured trap stiffness for the ASR metamaterial with various nanotweezer systems reported in the literature\cite{R51,R35,R14,R52,R27,R11}. 

\begin{table}[]
\centering
\caption{\textbf{Trap stiffness for nanoscale optical tweezers}}
\label{tab:my-table}
\resizebox{\textwidth}{!}{%
\begin{threeparttable}[b]
\begin{tabular}{cccccc}
\hline \hline
\textbf{Structure} & \multicolumn{1}{l}{\textbf{Method}} & \textbf{\begin{tabular}[c]{@{}c@{}}Trapping\\ Wavelength\\ (nm)\end{tabular}} & \textbf{\begin{tabular}[c]{@{}c@{}}Nanoparticle\\ (nm)\end{tabular}} & \textbf{\begin{tabular}[c]{@{}c@{}}Trap Stiffness\\ (fN/nm) to\\ 1mW/$\mu$m$^{2}$\end{tabular}} & \textbf{\begin{tabular}[c]{@{}c@{}}Scaled Trap Stiffness\tnote{a}\\ (fN/nm)  \end{tabular}} \\
\hline \hline
\begin{tabular}[c]{@{}c@{}}Silicon Nitride Photonic Crystal\\ [6pt] 
   \end{tabular} & exp.\tnote{b} & 1064 & 22 PS & 0.14 & 0.11 \\ 
\begin{tabular}[c]{@{}c@{}}Silicon Nano-antennas\\ [6pt] \end{tabular} & exp. & 1064 & 20 PS & 0.01 & 0.01 \\
\begin{tabular}[c]{@{}c@{}}Double Nanohole \\ [6pt] \end{tabular} & exp. & 820 & 20 PS & 0.10 & 0.10 \\
\begin{tabular}[c]{@{}c@{}}Rectangular Nanocavity \\ [6pt] \end{tabular} & sim.\tnote{c} & 1064 & 20 PS & 0.33 & 0.33 \\
\begin{tabular}[c]{@{}c@{}}Coaxial Nano-apertures \\ [6pt] \end{tabular} & sim. & 692 & 10 (n = 2\tnote{d}) & 0.19 & 1.56 (0.36\tnote{e}) \\
\begin{tabular}[c]{@{}c@{}}Connected Nano-aperture Array \\ [6pt] \end{tabular} & exp. & 980 & 30 PS & 0.84 & 0.25 \\
\multicolumn{1}{l}{Asymmetric Split-Rings Metamaterial} \\ (this work) & exp. & 930 & 20 PS & 8.65 & 8.65 \\ 
\hline \hline
\end{tabular}%
\begin{tablenotes}
     \item[a] $(x 10 nm/r)^{3}$, where \textit{r} is the radius of the particle.
     \item[b] experimental.
     \item[c] simulation.
     \item[d] an intermediate value between the refractive index of biomolecules and quantum dots\cite{R27}.
     \item[e] scaled by the Clasusius-Mossotti factor to the refractive index of polystyrene.
   \end{tablenotes}
\end{threeparttable}
}
\end{table}

In the last column, trap stiffness has been scaled to a particle with a 10-nm radius, since the stiffness is proportional to the particle volume \cite{R55}, and to the Claussius-Massoti factor for the refractive index in the case of the coaxial nano-aperture. The coaxial nano-aperture width is comparable to the width of our ASR metamaterial device. Based on Table 1, ASR metamaterial structures demonstrate a scaled trap stiffness approximately 86 times larger than that for a double nanohole and $\sim$ 24 times larger than that for a coaxial nano-aperture cavity. Moreover, we also compared our measured trap stiffness to those obtained for conventional optical tweezers\cite{R56}, using the same scaling methods without the contribution of Faxen's correction factor (see S3-SI). In this case, the ASR metamaterial has a scaled trap stiffness of 4.86 fN/nm/mW for a 20-nm PS particle; this is 180 times larger than the stiffness of 0.027 fN/nm/mW obtained for a 220-nm particle\cite{R56} using conventional optical tweezers. Considering that the particle used in the conventional trapping scheme is 11 times larger than the particle used in our work, we conclude that the FAPOT is about $10^{5}$ times more efficient than conventional optical trapping. 

Although trap stiffness does not provide direct information about the trapping force or the depth of the trapping potential, it does contain information about fluctuations in particle position and the viscous drag of the particle in the surrounding medium\cite{R56}. Recently, a plasmonic nanopore biosensor\cite{R57} was reported to hold and detect single molecules for extended trapping times in order to investigate perturbation of the molecule at hotspots. Even if this approach provides insights for precise nanotrapping optimization, the lack of augmented motion control of the biomolecule remains a challenge due to the weak trapping\cite{R57}. Therefore, a particularly stable trapping environment at low incident trapping laser intensity is mandatory to trap nanosized particles for extended measurement times. The high trap stiffness achieved in this work indicates the ability of our system to grab the particle efficiently, to pull it toward the hotspot, and to retain it there for a long time in order to study its structural characteristics and native dynamics. Therefore, our experimental trap stiffness magnitude is of fundamental importance for future understanding and exploitation of transition paths\cite{R4}, as well as investigations into the translocation mechanisms of biomolecules\cite{R70}. 

To understand the origin of trap stiffness enhancement in our experiments, we assume that the metamaterial creates an environment equivalent to a microcavity and we calculate the Purcell factor using the following equation\cite{R31}: 

\begin{equation}
    F_{p}=\frac{3}{4\pi^{2}}\left(\frac{\lambda_{p}}{n}\right)^{3}\frac{Q}{V_{mod}},
\end{equation}
where \textit{Q} is the quality factor (see Eq. S2-SI), \textit{$V_{mod}$} is the mode volume (see Eq. S3-SI), $\lambda_{p}=892$~nm is the peak wavelength of the transmission spectrum in Figure 1b, and $n = 1.33$ is the refractive index of the surrounding medium. 
Thence, we obtain a Purcell factor, \textit{$F_{p}$} = 184. Based on the observed enhancement of trap stiffness, \textit{i.e.}, \textbf{$\mu$} = 62.4 (Figure 4b), we calculate that the Purcell factor in our experiments, \textit{$F_{p,exp}$} = 198 (see SI-S4). Notably, Purcell factor equations used in this work approximate \cite{R31} the trap stiffness enhancement in our plasmonic tweezer platform. 

 Based on the above analysis, we assume that the powerful trapping achieved by our device originates from strong coupling between the Fano resonance and the trapped particle, as well as the ultra-small mode volume that enhances the density of photon flux through the nano-aperture. Indeed, the reasonable agreement between the experimentally observed Purcell factor of 198 and the theoretically calculated value of 184, strengthens our assumption. In addition, the small discrepancy between the values of the Purcell factor could result from the influence of the self-induced back action (SIBA) effect on the Fano resonance frequency. We attribute this behavior to the fact that, as the nanoparticle approaches the metamaterial to the point of local maximum field intensity, the motion of the nanoparticle changes the intracavity profile\cite{R85}. This may lead to an increase in the Purcell factor obtained from the trap stiffness enhancement. Plasmonic nano-aperture cavity trapping is believed to benefit from the SIBA effect, where the presence of a particle in the trap shifts the resonance peak close to the excitation wavelength, thereby creating a strong trapping force\cite{R17}. In our trapping experiments, we noticed a red-detuned resonance behavior. However, this consideration is beyond the scope of this current work and it will be a subject of future studies. 
 
Furthermore, in slow light devices such as photonic crystals\cite{R83,R84} and metamaterials\cite{R90} a decrease in the group velocity of the light corresponds to an increase in the  group refractive index, \textit{$n_{g}$}, which can lead to higher laser intensities. Particularly, a high \textit{Q}-factor implies that the light is stored for a few cycles in the metamaterial before being released; hence, propagation appears slower and the refractive index appears higher. Similarly, we assume that as the particle moves closer to the ASR metamaterial, this effectively increases the ambient group refractive index, leading to an increase in laser intensity. This behavior is directly related to the figure-of-merit (FOM). A nanoscale cavity is typically characterized by the FOM, which is proportional to the ratio of the \textit{Q}-factor to the cubic power of the ambient refractive index, \textit{i.e.}, \textit{Q/$n^{3}$}. Thereby, the FOM may decrease with an increasing refractive index. Considering the above analysis, we assume that the decreased trap stiffness enhancement factor at high trapping laser intensities (Figure 4b), may be due to a decrease in the FOM. 

Finally, the heating effect arising from gold's absorption of the laser light can increase the Brownian motion of the trapped particle, thereby affecting the trap stiffness\cite{R69}. We observed a deviation of the linear behavior (Figures 3 and 4b) that may be due to a weak thermal effect. It is generally believed that the temperature increase for nano-apertures\cite{R69,R35} remains below a few Kelvin and that the metal layer acts as an efficient heat sink to further dissipate Joule heating. Therefore, based on the literature, we assume that the thermal-induced fluid convection is weak\cite{R69,R35} and makes a negligible contribution to the trapping process.

\section{Conclusions}

An array of asymmetric split-rings 
with the ability to support a Fano resonance peak 
and boost near-field enhancement was fabricated in order to demonstrate a new platform for the next generation of optical trapping instruments. Trapping performance was investigated as a function of incident laser intensity and trapping wavelength. The scaled trapping efficiency of 8.65 fN/nm/mW for 20-nm polystyrene particles using a 930-nm trapping laser is the highest value reported for dielectric nanoparticle trapping. The mechanism underpinning the trap stiffness enhancement results from the ultra-small effective electromagnetic mode volume and the Purcell effect contribution. Moreover, a small contribution of the modifications of the Fano resonance wavelength due to the nano-aperture effect was observed. Our approach can be used to qualitatively understand the folding mechanism of biomolecules, such as proteins and nucleic acids, that is related to the evolution of various diseases and which could contribute to drug discoveries.


\subsection{Corresponding author}
\textbf{Email:} sile.nicchormaic@oist.jp

Domna G. Kotsifaki $^{}$\orcidA{}:0000-0002-2023-8345

Viet Giang Truong $^{}$\orcidB{}:0000-0003-3589-7850 

S\'ile Nic Chormaic $^{}$\orcidC{}:0000-0003-4276-2014

\subsection{Author Contributions}

SNC and VGT conceived the experiments. DGK and VGT performed the experiments and analyzed the data. DGK wrote the manuscript. All authors discussed the results and contributed to the manuscript.


\subsection{Notes}
The authors declare no competing financial interests.

\begin{acknowledgement}
This work was supported by funding from Okinawa Institute of Science and Technology Graduate University. The authors are grateful to Nikitas Papasimakis for useful comments on the manuscript. The authors would like to thank Simon Peter Mekhail and Metin Ozer for technical assistance and Emi Nakamura for general research support.

\end{acknowledgement}

\bibliography{achemso-demo}

\newpage

\begin{suppinfo}

\subsection{S1 Experimental Set-up}

\begin{figure}
\setcounter{figure}{0}
\makeatletter 
\renewcommand{\thefigure}{S\arabic{figure}}
\makeatother
\centering
\includegraphics[width = 150 mm, height = 140 mm]{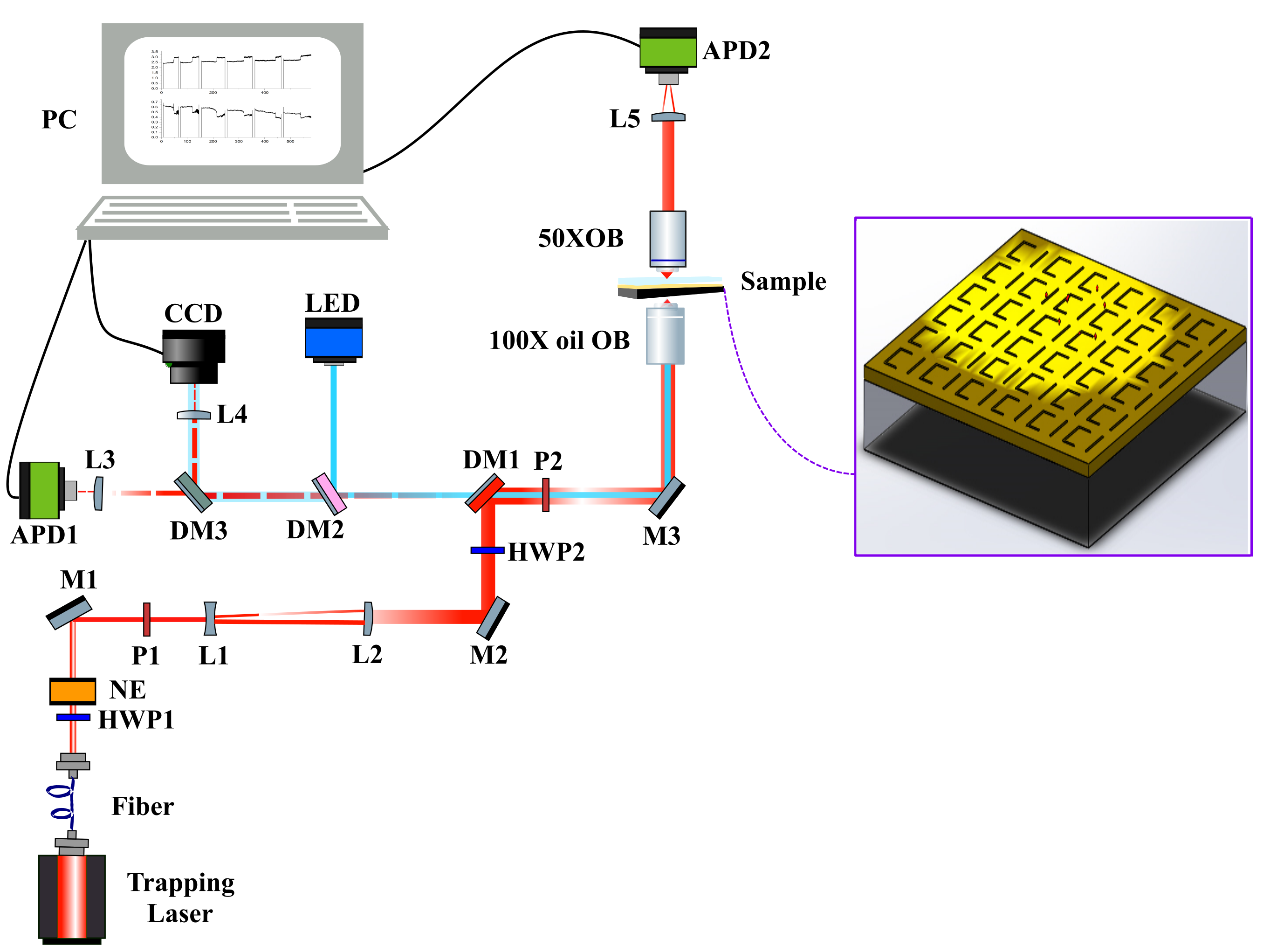}
\caption{Schematic diagram of the optical trapping set-up.The rectangular inset shows a schematic of the metamaterial device. NE: noise eater (NEL03/M Thorlabs); M: mirror; L: lens; HWP: half-wave plate; OB: objective lens; P: polarizer; APD: avalanche photodiode; LED: illumination source; CCD: camera; DM: dichroic mirror.}
\label{FigureS1}
\end{figure}
\newpage
\subsection{S2 Multiple Nanoparticle Trapping}

\begin{figure}
\setcounter{figure}{1}
\centering
\makeatletter 
\renewcommand{\thefigure}{S\arabic{figure}}
\makeatother
\centering
\includegraphics[width = 160 mm, height = 95 mm]{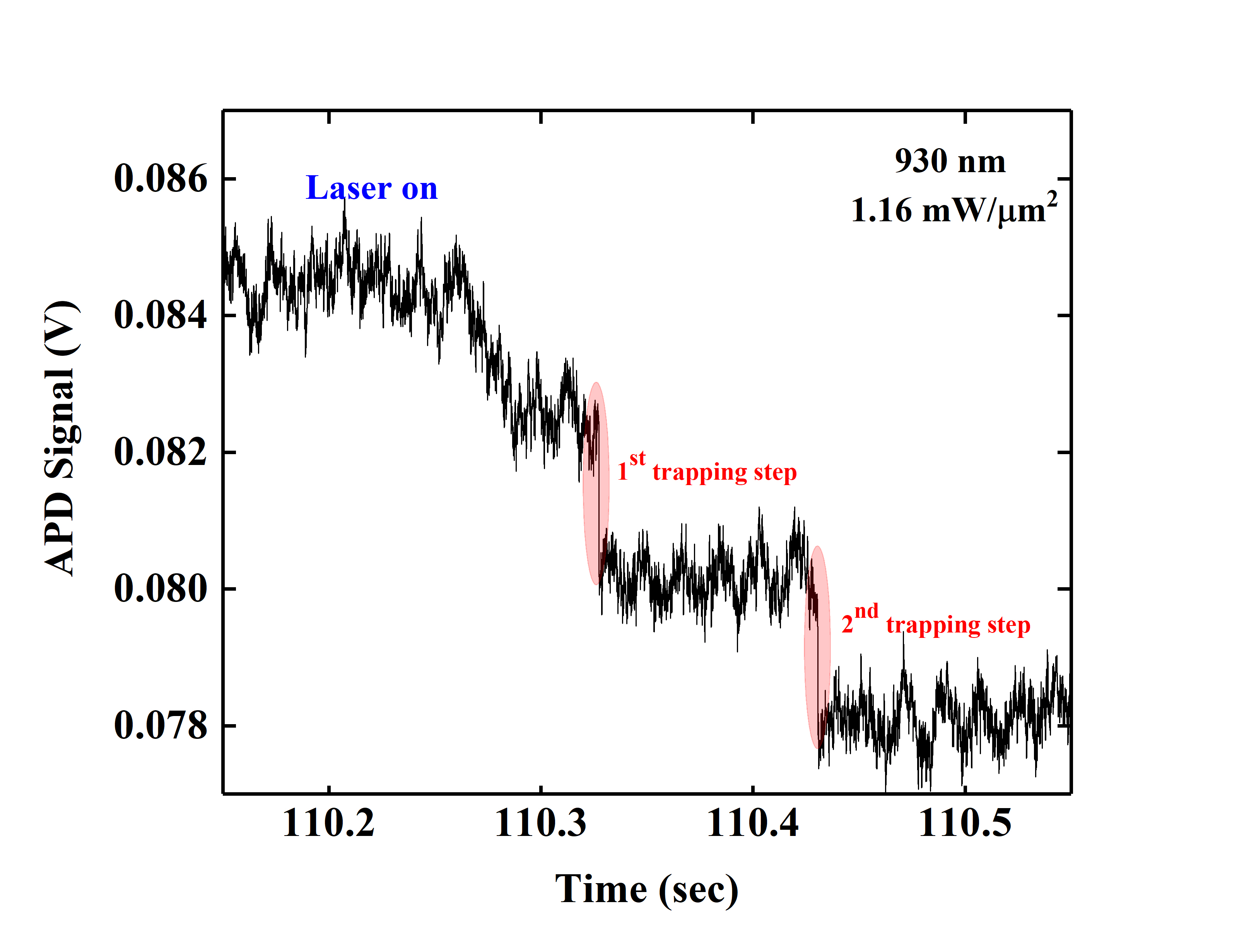}
\caption{Time trace of the APD signal through the ASR metamaterial in heavy water containing suspended 20-nm PS particles as a function of the trapping laser beam. Gray regions indicate times when the laser was blocked. Multiple steps in the APD signal confirm the ability of the FAPOT to activate multiple trapping sites (red regions).}
\label{FigureS2}
\end{figure}

\subsection{S3 Fax\'{e}n's Correction}
The Faxen correction of Stokes' law can be obtained from the equation in Ref. [S1]:

\setcounter{equation}{0}
\def\theequation{S\arabic{equation}}
\begin{equation}
\gamma=\frac{6\pi\eta r}{[1-\frac{9}{16}(\frac{r}{h})+\frac{1}{8}(\frac{r}{h})^{3}-\frac{45}{256}(\frac{r}{h})^4-\frac{1}{16}(\frac{r}{h})^5]},
 \end{equation}
where $\eta$ is the viscosity of the medium,  \textit{i.e.}, heavy water in our case, \textit{r} is the radius of the PS particle,  and \textit{h} is the distance from the particle to the ASR wall.

\subsection{S4 Numerical Calculations}

The quality factor of a Fano resonance can be obtained from the formula in Ref. [S2]:
\setcounter{equation}{1}
\def\theequation{S\arabic{equation}}
\begin{equation}
  Q=\frac{\lambda_{p}+\lambda_{d}}{|\lambda_{p}-\lambda_{d}|},
\end{equation}
where \textit{$\lambda_{d}$} is the wavelength of the dip in the transmission spectrum in Figure 1b (\textit{$\lambda_{d}$} = 983 nm).
The mode volume is calculated from Eq. S3, where \textit{$\alpha$} = 310~nm, \textit{t} = 164.4~nm, \textit{w} = 44.3~nm,  and the thickness of the gold film in which the mode is assumed to be confined is \textit{h} = 50~nm), see Ref. [S3]:
\begin{equation}
    V_{mod}=2(\alpha+t)wh. 
\end{equation}

\noindent The Purcell factor for the above-mentioned parameters is \textit{$F_{p}$} = 184. Additionally, we observed an enhancement to trap stiffness, \textbf{$\mu$} = 62.4 (Figure 4b). In our case, the FWHM of the trap potential along the \textit{x}-direction  is  \textit{p} = 50~nm, theoretically. Consequently, the epxerimentally obtained  Purcell factor is determined using  Eq. S4 of Ref. [S3]:

\begin{equation}
    F_{p,exp}=\frac{\mu D^{2}p}{V_{mod}}=198.
\end{equation}


\subsection{Supporting Information References 
}

(S1) Goldman, A. J.; Cox, R. G.; Brenner, H. Slow viscous motion of a sphere parallel to a plane wall—I Motion through a quiescent fluid. \textit{Chem. Eng. Sci.} \textbf{1967}, \textit{22}, 637–651.

\noindent (S2)Fan, Y.; Zhang, F.; Shen, N.-H.; Fu, Q.; Wei, Z.; Li, H.; Soukoulis, C. M. Achieving a high-Q response in metamaterials by manipulating the toroidal excitations. \textit{Phys. Rev. A} \textbf{2018}, \textit{97}, 033816.

\noindent (S3) Tanaka, K.; Plum, E.; Ou, J. Y.; Uchino, T.; Zheludev, N. I. Multifold enhancement of quantum dot luminescence in plasmonic metamaterials. \textit{Phys. Rev. Lett.}, \textbf{2010}, \textit{105}, 227403.

\end{suppinfo}







\end{document}